\documentclass[useAMS]{mn2e}
\usepackage{lscape}
\usepackage{rotating}
\usepackage{amssymb}
\def\tevo{\hbox{t$_{\rm evo}$}}

\def\msun{\hbox{M$_\odot$}}

\def\msunyr{\hbox{M$_\odot$yr$^{-1}$}}

\def\cm3{\hbox{cm$^{-3}$}}
\def\hoii{\hbox{Holmerg~{\sc II}}}
\let\la=\lesssim
\let\ga=\gtrsim 
\input psfig.sty
\input epsf.sty
\usepackage{graphicx}
\usepackage{epsfig}
\title[The evolution of stellar structures]
{The evolution of stellar structures in dwarf galaxies}
\author[Bastian et al.] {N. Bastian$^{1}$, D.R. Weisz$^{2}$,  E.D. Skillman$^{2}$, K.B.W. McQuinn$^{2}$, A.E. Dolphin$^{3}$,\newauthor  R.A. Gutermuth$^{5,6}$, J.M. Cannon$^{4}$, B. Ercolano$^{1}$, M. Gieles$^{7}$, R.C. Kennicutt$^{7}$, \newauthor F. Walter$^{8}$ \\
$^1$ School of Physics, University of Exeter, Stocker Road, Exeter EX4 4QL, UK\\
$^2$ Astronomy Department, University of Minnesota, Minneapolis, MN 55455, USA\\
$^3$ Raytheon Company, 1151 E Hermans Rd, Tucson, AZ 85756 \\
$^4$ Department of Physics \& Astronomy, Macalester College, 1600 Grand Avenue, St. Paul, MN 55125\\
$^5$ Five College Astronomy Department, Smith College, Northampton, Mass.  01063  USA\\
$^6$ Department of Astronomy, University of Massachusetts, Amherst, Mass.  01002  USA\\
$^7$ Institute of Astronomy, University of Cambridge, Madingley Road, Cambridge CB3 0HA, UK\\
$^8$ Max-Planck-Institut f\"{u}r Astronomie, K\"{o}nigstuhl 17, D-69117 Heidelberg, Germany\\
}
\date{Accepted. Received; in original form}
\pagerange{\pageref{firstpage}--\pageref{lastpage}}
\pubyear{2009}
\begin{document}
\maketitle
\label{firstpage}
\begin{abstract}
%\vskip 70truemm

We present a study of the variation of spatial structure of stellar populations within dwarf galaxies as a function of the population age.  We use deep {\it Hubble Space Telescope/Advanced Camera for Surveys} imaging of nearby dwarf galaxies in order to resolve individual stars and create composite colour-magnitude diagrams (CMDs) for each galaxy.  Using the obtained CMDs, we select Blue Helium Burning stars (BHeBs), which can be unambiguously age-dated by comparing the absolute magnitude of individual stars with stellar isochrones.  Additionally, we select a very young ($\lesssim10$~Myr) population of OB stars for a subset of the galaxies based on the tip of the young main-sequence.  By selecting stars in different age ranges we can then study how the spatial distribution of these stars evolves with time.  We find, in agreement with previous studies, that stars are born within galaxies with a high degree of substructure which is made up of a continuous distribution of clusters, groups and associations from parsec to hundreds of parsec scales.  These structures disperse on timescales of tens to hundreds of Myr, which we quantify using the two-point correlation function and the $Q$-parameter developed by Cartwright \& Whitworth (2004).  On galactic scales, we can place {\it lower limits} on the time it takes to remove the original structure (i.e., structure survives for at least this long), \tevo, which varies between $\sim100$~Myr (NGC~2366) and $\sim350$~Myr (DDO~165).  This is similar to what we have found previously for the SMC ($\sim80$~Myr) and the LMC ($\sim175$~Myr).  We do not find any strong correlations between \tevo\ and the luminosity of the host galaxy. 

\end{abstract}
\begin{keywords} galaxies - dwarf; galaxies - individual (DDO~165, Holmberg~{\sc II}, IC~2574, NGC~784, NGC~2366, NGC~4068)
\end{keywords}

\section{Introduction}

Young stars are observed to be born in a clustered, hierarchical or fractal distribution within galaxies, and the goal of the present work is to measure the timescale over which stellar structures in dwarf galaxies disperse. In Gieles et al.~(2008; hereafter G08) and Bastian et al.~(2009; hereafter B09) we investigated the stellar distribution in the SMC and LMC, respectively. In these works we developed a series of statistical techniques with which structures can be quantified and their evolution traced as a function of time. We found that for the SMC and LMC stellar substructure was erased gradually, being indistinguishable from the background distribution by $75$~Myr and $175$~Myr, respectively. These timescales are comparable to the crossing time (t$_{\rm cross} = R_{\rm galaxy}/\sigma$ where $\sigma$ is the stellar velocity disperison) of the galaxy which led to the conclusion that general galactic dynamics were responsible for the removal of structure, rather than local effects such as Ópopping star clustersÓ (also known as infant mortality, i.e., cluster expansion due to the loss of residual gas left over from the star-formation process - Boily \& Kroupa~2003; Bastian \& Goodwin~2006).

One of the major limitations of the previous works was in defining samples of nearly coeval stars (i.e. stars of similar age). Previously, we selected groups of stars from regions in the colour-magnitude diagram along the main-sequence (MS) and then carried out simulations to estimate the mean age and spread of stars within that region of space in the colour-magnitude diagram (CMD). This resulted in a nearly coeval population only for the youngest ages, whereas regions further down the CMD were made up of older stars as well as some contribution of younger stars. Our results were tested against the evolution of the distribution of stellar clusters with known ages, and overall we found good agreement.

The post-MS evolutionary phase of young ($\lesssim$1 Gyr) intermediate mass  ($\gtrsim$ 3 \msun) stars provides an alternative method of age dating that avoids the age degeneracies of MS stars.  Of particular importance are core blue helium burning stars (BHeBs) which have a one-to-one relationship between age and luminosity (e.g.,~Bertelli et al.~1994), making it possible to assign ages to individual BHeBs on an optical CMD. The short lifetimes of BHeBs, $\sim$ 10\% of their MS lifetimes (e.g.,~Chiosi, Bertelli, \& Bressan~1992), allow for a clear identification of coeval populations with minimal confusion between stars of different ages (Dohm-Palmer et al.~1997).  For a given age, BHeBs are $\sim$ 2 magnitudes brighter compared to MS stars, thus providing higher photometric precision and longer look back times ($\sim$ 500 Myr;  e.g.,~Dohm-Palmer et al.~1997, Weisz et al.~2008).  Although there are fewer BHeBs than MS stars, the precise age-dating, possible with BHeBs, compensates for the reduction in numbers and provides a more secure final sample of stars.  In those cases where the photometry is sufficiently deep, the star formation histories derived from the BHeB stars are consistent with  those derived from the main sequence stars (e.g., Dohm-Palmer et al.~2002).

Imaging of resolved stellar populations, specifically BHeBs, in nearby star forming dwarf galaxies, i.e., dwarf irregular galaxies (dIrrs), could reveal fundamental processes driving the evolution of stellar structures.  Unlike more massive galaxies, dIrrs are close to solid body rotators (e.g.,~Skillman et al.~1988, Skillman~1996), implying that the locations of stars are not significantly disturbed by external mechanisms such as sheer or spiral density waves.   Combined with precise ages of BHeBs, the dynamic simplicity of dIrrs may provide new insight into the intrinsic evolution of stellar structures.

Ideally, one would like to study not only the timescale over which stellar structures are erased, but also how this depends on spatial scale.  Sub-cluster  ($<1$~pc) structures are erased quickly (Goodwin \& Cartwright~2004) while larger structures take more time to disperse.  The exact dependence of the structural evolutionary timescale and the spatial scale probed will be essential in confirming the cause of the evolution.  

There have been many recent studies on quantifying the distribution of point-like (i.e., stellar) structures on a variety of physical scales. Near- and mid-IR studies of star-forming regions on scales of  $0.1- 10$~pc have revealed a wealth of structural information (Lada \& Lada 2003; Cartwright \& Whitworth 2004; Gutermuth et al. 2005, 2008, 2009; Schmeja \& Klessen 2006; Koenig et al. 2008, Kraus \& Hillebrand 2008) with many regions appearing centrally concentrated and relaxed while others are still largely fragmentary. On intermediate spatial scales ($1 - 100$pc) extragalactic studies on objects like NGC 346 in the SMC have revealed that star-forming regions are made up of many individual clusters and also distributed star-formation (e.g., Sabbi et al. 2007; Schmeja, Gouliermis, \& Klessen 2009). On galactic scales, it appears that there is not a characteristic size for star- forming regions, and that there exists a continuous hierarchy of structure from pc to kpc scales (Elmegreen \& Elmegreen~2001, Elmegreen et al.~2006; Bastian et al. 2005, 2007, 2009; G08; Odekon 2008). The framework that is developing is that stars form in a highly structured distribution on all scales probed so far, and that these structures are erased on approximately the crossing time, regardless of scale.

The number of nearby dwarf galaxies with high quality optical imaging available has significantly increased in the past decade, thanks to a number of Hubble Space Telescope survey programs (e.g.,~Tully et al.~2006, Weisz et al.~2008, Dalcanton et al.~2009).  The excellent spatial resolution afforded by HST allows for precise photometry and age dating of individual BHeBs.  In this paper, we have selected six star forming dwarf galaxies in the Local Volume (Weisz et al.~2008, McQuinn et al.~2010) in which we will explore the evolution of stellar structure using a statistically significant number of BHeBs.  

This paper is organised as follows.  In \S~\ref{sec:obs} we introduce the observations used in the present study.  The technique of age-dating BHeB stars is discussed in \S~\ref{sec:age_dating} and in \S~\ref{sec:structure} we introduce the methods used to quantify the evolution of spatial structures along with their caveats.  The results for each individual galaxy are given in \S~\ref{sec:results} and in \S~\ref{sec:discussion} we summarise and discuss our main findings.

\section{Observations and Photometry}
\label{sec:obs}

Observations of all galaxies were taken with the Advanced Camera for Surveys (ACS, Ford et al.~1998) aboard HST.  DDO~165, Holmberg {\sc II}, NGC~2366, and IC~2574 were observed in the F555W and F814W filters as part of a survey of the M81 Group dwarf galaxies (GO-10605 - Weisz et al.~2008; also GO-9755, PI Walter, was used for IC~2574), while NGC~4068 (GO-9771 - Karachentsev \& Kashibadze~2006) and NGC~784 (GO-10210 - Tully et al.~2006) were both observed in the F606W and F814W filters.   The data we use in this paper have been processed as described in Weisz et al.~(2008) and McQuinn et al.~(2010).  Here, we briefly summarize the data reduction process.

Images were all downloaded from the HST archive following standard reduction and processing with the HST pipeline. To ensure uniformity among our data, photometry of all images was performed using DOLPHOT\footnote[1]{http://purcell.as.arizona.edu/dolphot/}, a version of HSTPHOT optimized for the ACS (Dolphin~2000).  Cosmic rays, hot pixel residuals, and extended objects were all rejected based on their brightness profiles.  The remaining objects were subjected to further photometric cuts.  Specifically, we considered well measured stars to have a signal-to-noise ratio $>$ 5, $| sharp_{1} + sharp_{2} | < 0.27$, and $ crowd_{1} + crowd_{2}  < 1.0$.  The definitions of these photometric criteria can be found in Dolphin~(2000) and Dalcanton et al.~(2009).  Full descriptions of data reduction for NGC~4068 and NGC~784 can be found in McQuinn et al.~(2010), while identical details for DDO~165, Holmberg {\sc II}, NGC~2366, and IC~2574 are presented in Weisz et al.~(2008).

Galaxies considered in this paper are a subset of $\sim$ 25 star forming dwarf galaxies presented in Weisz et al.~(2008) and McQuinn et al.~(2010).  From this larger sample, we first eliminated galaxies whose CMDs did not have a sufficient population of BHeBs.  From the remaining sub-sample, we then considered only galaxies which have clearly separable MS and BHeB sequences.  Simulated CMDs (Gallart, Zoccali, \& Aparicio~2005, Weisz et al.~2008) show that the MS and BHeBs are typically separated by several tenths of a magnitude in colour, enough to trivially identify them as two separate sequences.  However, in the event of intense periods of SF or `starbursts', a feature which describes many of the galaxies in the initial sample of $\sim$ 25 (McQuinn et al.~2010), internal extinction effects can smear out the separation between the MS and BHeBs (e.g.,Cannon et al.~2003, McQuinn et al.~2009, 2010).  Additionally, higher values of metallicity lead to higher dust-to-gas ratios with greater concomitant differential extinction. This effect can make it difficult to discern the difference between individual young MS and BHeB stars.  As discussed in \S~\ref{sec:caveats} and Appendix~B in B09, inclusion of young MS stars in the BHeB sample can introduce false structure detections, potentially biasing the results.  Thus, we eliminated galaxies without clearly separated MS and BHeB sequences.  Finally, we attempted to include larger, actively star-forming galaxies to our sample, in particular NGC~4449 and NGC~5253.  Unfortunately, these had to be eliminated due to completeness issues in the high surface brightness central regions of the galaxies for the older (fainter) BHeB stars.  Application of these criteria to the initial sample resulted in the six galaxies we consider for this study.

\begin{figure}
\includegraphics[width=8cm, angle=0]{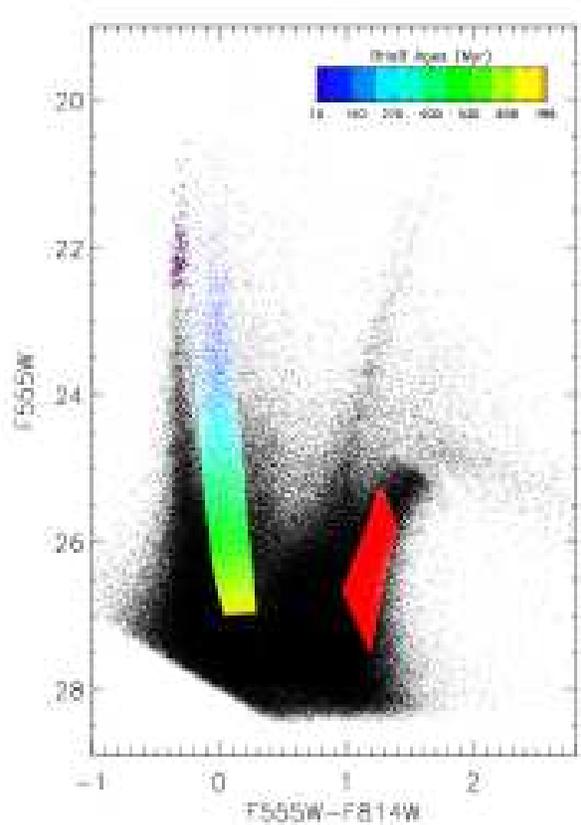}
\caption{The colour-magnitude diagram of stars in Holmerg~II.  The magenta filled circles at the tip of the main sequence represent OB stars used in the analysis.  The coloured dots (from blue to orange) represent BHeB stars used in this study, colour coded to reflect their age (quantified in the colour bar).  The red region denotes RGB stars used as the background template population in the two-point correlation function analysis (see \S\S~\ref{sec:ref_pop} \& \ref{sec:tpcf}).}
\label{fig:cmd_hoii}
\end{figure}

\begin{figure*}
\includegraphics[width=14cm, angle=0]{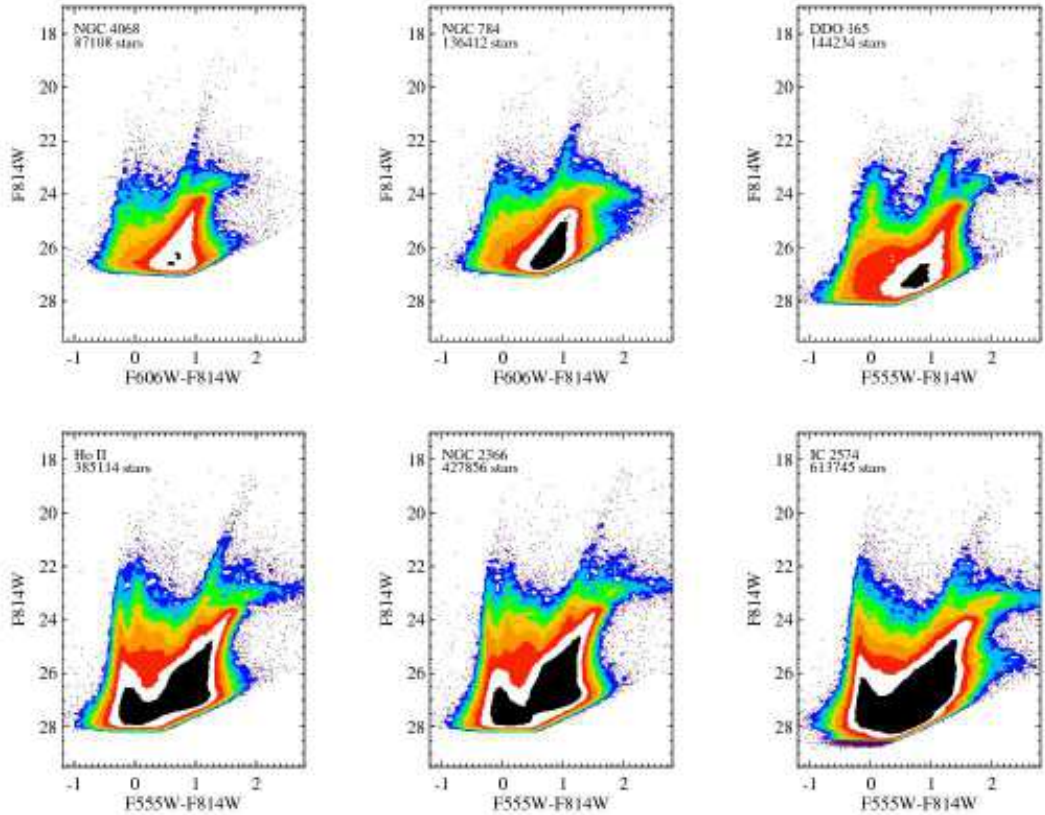}
\caption{The observed colour-magnitude diagrams (CMDs) of the galaxies in the present study.}
\label{fig:cmds}
\end{figure*}

The CMD of Holmerg~II is shown in Fig.~\ref{fig:cmd_hoii} and its main features are indicated.  Additionally, the observed CMDs of all galaxies used in the present student are shown in Fig.~\ref{fig:cmds}.

\section{Age dating Method}
\label{sec:age_dating}

Precise age dating is essential to understanding the evolution of stellar structures.  Fortunately, luminous BHeBs have high degrees of photometric accuracy.  Combined with the one-to-one age luminosity relationship predicted by stellar models (e.g., Bertelli et al.~1994), it is possible to assign unambiguous ages to BHeBs $\lesssim$ 500 Myr old (e.g.,~Dohm-Palmer et al.~1997).  In this paper, we use the stellar models of Marigo et al.~(2008), which include the massive star models of Bertelli et al.~(1994) transformed into ACS instrumental magnitudes.

For our choice of observational wavelengths, the BHeB sequence is a primarily vertical sequence of stars located redward of the young MS, which generally occupies a relatively unique position on the CMD (see Fig.~\ref{fig:cmd_hoii}).  The exact location and shape of the BHeB sequence depends upon both metallicity and differential foreground and/or internal extinction effects.  The former primarily influences the colour of the locus of BHeB stars, while the latter can introduce broadening of the BHeB and MS tracks, possibly blending the two together in instances of high extinction (e.g.,~Cannon et al.~2003).  The galaxies we consider in this paper have been selected in part because they have low values of differential extinction, which makes for a clear separation between the MS and BHeBs.  

The distinct separation between these sequences makes it possible to visually select the BHeBs.  We select a polygonal region around the BHeBs, with the blue and red colour limits selected to tightly enclose the BHeB sequence (see Fig.~\ref{fig:cmd_hoii}).  The bright end of the BHeBs is determined using the 10 Myr MS turn off magnitude.  For ages $<10$~Myr, BHeB and MS stars are usually sparse and
 blend together making the separation of sequences not as clear.  The faint end of the BHeB sequence extends into the bright end of the red clump (RC).  The precise age and magnitude of where BHeBs intersect the RC depends upon physical parameters such as metallicity and SFH (which can affect the distribution of RC stars) as well as the filters used for the observations.  
 The faint limit of the BHeB is difficult to identify visually.  We thus use simulated CMDs to determine the magnitude limit at which the BHeBs merge into the RC.  Inclusion of some RC stars among our oldest BHeBs stars is still likely.  We note, however, that RC stars generally have a smooth spatial distribution, and thus their inclusion as BHeB stars serves to wash out, and not enhance, signals of stellar structure.  

Comparison of the observed BHeB sequence with stellar models allows us to assign ages to individual BHeBs.  To convert from magnitude to age, we first select the appropriate metallicity isochrone based on H{\sc ii} region abundances, if available.  Otherwise we use the best fit isochronal metallicity as determined by quantitative derivation of the SFH (e.g.,~Dolphin~2002).   The stellar models contain the ages of the BHeBs sequence tabulated at time intervals of $\log(age)$ $=$ 0.05.  However, our statistical structure analysis is driven by having an approximately equal number of stars per time bin, and thus we seek a continuous age distribution for individual stars, which permits a flexible choice in age binning.  To create a continuous age distribution, we consider the BHeB sequence in $\log(age)$--absolute magnitude space.  Here, we can functionally approximate the BHeB sequence by a 5th order polynomial, which provides for a one-to-one analytic relation between age and luminosity.  From this relationship we can convert the obseved magnitudes of BHeBs into ages for individual stars.  

Errors on the ages are derived from the photometric uncertainties of the individual stars.  The bright and faint photometric errors for each star are propagated through the analytic age--luminosity relationship, and are translated into age error estimates.  For the galaxies considered in this paper, the uncertainties in the assigned ages are typically $\sim$ 10\%.  This error process does not account for systematic uncertainties inherent to the choice of isochrones.  Thus, while the resultant relative ages are robust, the absolute ages could depend on the choice of stellar model.  We note that when comparing the Geneva and Padova evolutionary models for age assignment to BHeBs, Dohm-Palmer et al.~(1998) find that the two models provide nearly identical results.

While BHeBs allow us to trace structure from $\sim$ 10-500 Myr ago, we can also use young MS stars for additional structural information.  We select MS stars that are brighter than the 10~Myr MS turnoff magnitude and bluer than the blue edge of the selected BHeBs.  Due to the brightening of MS stars during their evolution, some older stars will fall into our selected region of colour/magnitude space.  From Figure~4 in Weisz et al.~(2008) we see that stars as old as 25~Myr may enter our selection, but that they are rare, and that the sample is dominated by younger stars.  From the simulations discussed in \S~\ref{sec:example} we assign a representative age of 8~Myr for stars that pass this selection, and will refer to them as OB stars, although we note that our selection does not include all true OB stars in each galaxy.  

\subsection{Reference Population Selection}
\label{sec:ref_pop}

For a comparison reference population, we selected red giant branch (RGB) stars from the CMDs. The RGB stars represent an older stellar component that has had sufficient time to diffuse and erase signs of local structure.  

The reference population was selected by drawing a box in the RGB at least 0.5 mags fainter than the tip of the red giant branch (TRGB).  The colour limits were restricted to be redder than the RC, to avoid possible selection of RHeBs, slightly bluer than the red edge of the RGB.  An example of the RGB stars selected in \hoii\ is show in Fig.~\ref{fig:cmd_hoii}.

\subsection{An Example of the Star Selection: Holmerg~{\sc II}}
\label{sec:example}

In this section, we describe the star selection method for the CMD of the M81 Group dwarf irregular galaxy \hoii.  The CMD presented in Fig.~\ref{fig:cmd_hoii} shows the MS, BHeB, and RGB stars used for tests of structure evolution in \hoii.

The selection of BHeBs requires setting both colour and magnitude limits around the BHeB sequence.  \hoii, like many low metallicity galaxies, shows a clear separation between the BHeBs and MS stars over several magnitudes ($m_{F555W}$ $\sim$ 21--25).  For slightly fainter magnitudes, the MS broadens due to photometric effects, and the BHeB sequence curves sightly redward. However, it is possible to visually identify the separate sequences and assign a blue colour limit.  The red colour limit is also determined by eye and is based upon the decreasing density of BHeBs as a function of increasing colour.  The luminous stars between the BHeB and red helium burning (RHeB) sequences are primarily HeBs transitioning between the two phases.  Luminosity evolution during this transition is roughly constant (Marigo et al. 2008), thus including them does not strongly affect the results.  Main sequence turn-off (MSTO) stars are known to also cross the BHeB sequence, however, the relative number of stars is only a few percent (Dohm-Palmer et al. 1997), and does not significantly affect the BHeB selection.

The BHeBs merge into the RC at $m_{F555W}$ $\sim$ 26.5.  Inclusion of some RC stars in the BHeB selection is probable, due to confusion at the faint limit.  However, RC stars are several Gyr in age, and do not exhibit clustered spatial structure.  Thus, if counted as BHeBs, they tend to smear out structure, rather than enhancing it.  To quantify the faint BHeB limit, we constructed a simulated CMD for a constant SFH, which was convolved with the \hoii\ artificial star tests to account for observational and completeness effects.  We then applied the same colour selection selection criteria to the simulated CMD, for which we know all stellar ages and masses, and determined the spread in BHeB age as a function of $m_{F555W}$.   Typically, the age spread at a given magnitude for BHeBs is $\sim$ 10\%.  However, at $m_{F555W}$ $\sim$ 27 ($\sim$ 550 Myr), the spread increases to 20\% and continues increasing thereafter.  Although we have selected BHeBs stars in Fig.~\ref{fig:cmd_hoii} older than this limit, we note the uncertainties at the oldest ages are as high as 40\%.

BHeBs are only reliable age indicators for ages $\gtrsim$ 10 Myr.  To gain additional information on structure in the most recent 10 Myr, we selected luminous MS stars using the isochrones of Marigo et al.~(2008).  The faint limit is based upon the 10 Myr MSTO isochrones, $M_{F555W}$ $\sim$ $-$4.  However, MS luminosity evolution causes slightly older stars to become brighter than this limit.  Although this is a rapid phase in a star's evolution, we chose a more conservative values of $M_{F555W}$ $=$ $-$5 ($m_{F555W}$ $=$ 22.65 - at the average distance to our targets) to minimise sources of contamination. The red colour limit of the OB stars was set by the blue edge of the selected BHeBs.  We also required MS stars to have a colour of $m_{F555W}$ $-$ $m_{F814W}$ $<$ 0.5.  There are $\sim$ 10 objects blueward of this limit, which could be binaries systems or unresolved blends, however they are not likely single OB type stars. 

We have also carried out consistency tests to check the number of observed vs. expected OB stars.  We simulated a CMD (30 realisations) with the star-formation history of \hoii\ based on the analysis of W08 for the past 10~Myr.  A SFR of 0.17~\msunyr\ was adopted for the past 4~Myr based on the H$\alpha$ flux and the SFR from 4 to 10~Myr was derived from the analysis of the BHeB stars (see W08).  We adopt a stellar IMF with index, -1.30 (i.e. close to the Salpeter value), from 0.15 to 100~\msun\, that covers the full range of masses for 10~Myr.  This yields $8.8*10^{5}$ stars, $160\pm12$ of which pass our colour/magnitude selection criteria of being a high-mass MS star.  The lowest mass star that passes our criteria has a present day mass of $\sim20$~\msun.  We observe 126 stars in our selection box indicating that not all high-mass stars in the galaxy made it through our selection process.  However, we only accepted 'point-like' objects in our analysis, so all stars in clusters or associations would have been removed from the sample.  Additionally, some stars may have been removed from the sample due to the complicated background found in H{\sc ii} regions.  BHeB stars, being older and more distributed on average, are not expected to suffer from these selection effects.  We conclude that we are picking up the majority of the expected MS massive stars.  The mean and median age of the stars meeting our criteria in selecting massive MS stars is $\sim7-8$~Myr.  We therefore adopt this age as a representative value for our MS population.

To provide a reference population, we sub-sampled the RGB using a simple polygon.  The bright and faint limits were selected to run parallel to the TRGB, while the red and blue limits were chosen to to minimise RHeB, Red Clump (RC), and Asymptotic Giant bBranch (AGB) star contamination.

The CMD in Fig.~\ref{fig:cmd_hoii} show the OB stars (purple points; 126 stars), BHeBs (age coloured sequence; 23,653 stars), and RGB reference population (red points; 28,919 stars).  The  corresponding spatial distribution of these populations are shown in Figure~\ref{fig:hoii_spatial}.  The position and derived age of all BHeBs and young stars is given in Table~\ref{table:stars}.   The error in age for each star is derived from its photometric error and does include uncertainties in the models used.

\begin{table}
\begin{center}
{\scriptsize
\parbox[b]{8cm}{
\centering
\caption[]{The position, age and estimated age error for all BHeB and young stars in our sample.    Stars that have a 'MS' after their ID number were selected as young main sequence stars (see \S~\ref{sec:age_dating}).  The full version is available in the electronic version of the paper.}
\begin{tabular}{c c c c c c }
\hline
\noalign{\smallskip}
Galaxy   &  star ID & x & y & age &  age error \\ % & median age & standard deviation in the age\\
               &            & (pixels)& (pixels)  & (Myr)    &   (Myr) \\ %& (Myr) & (Myr)\\ %($10^4 M_{\odot}$) & (mag)  & (mag) & (mag) \\

\hline
Holmberg II  &      1&  2052.95&  2274.90&    6.7&      0.2\\
Holmberg II  &      2&  2164.82&  1628.10&    7.2&      0.2\\
Holmberg II  &      3&  1459.65&  2061.96&    6.7&      0.2\\
Holmberg II  &      4&  2368.83&  1843.51&    7.7&      0.2\\
Holmberg II  &      5&  2858.23&  2619.56&    8.4&      0.2\\
Holmberg II  &      6&  2473.65&  2227.54&    8.1&      0.2\\
Holmberg II  &      7&  2750.72&  2117.37&    9.0&      0.2\\
Holmberg II  &      8&  2639.68&   624.23&    9.1&      0.2\\
Holmberg II  &      9&  1625.98&  1657.29&    9.4&      0.2\\
Holmberg II  &     10&  2752.44&  1774.87&    9.6&      0.2\\
\noalign{\smallskip}
\noalign{\smallskip}
\hline
\end{tabular}
\label{table:stars}
}
}
\end{center}
\end{table}

\section{Structural Analysis}
\label{sec:structure}

Once the age of each star has been estimated, we can begin to study their spatial distribution as a function of age. For this, we break the sample up into discrete age bins, and apply statistical tests to each distribution. We apply two tests to the data, each will be discussed in some detail below; 1) the two-point correlation function (TPCF) and 2) the Q- parameter (Cartwright \& Whitworth 2004).

In Fig.~\ref{fig:hoii_spatial} we show an example from the survey, \hoii.  In each panel we show the spatial distribution of 500 stars (except for the first panel where only 126 stars are used) stochastically sampled with the mean age given for four age bins selected from the BHeB sample, as well as for the OB stars and RGB star (reference) sample.  Examples for each galaxy are given in Appendix~\ref{sec:app1}.

\subsection{Two point correlation function (TPCF)}
\label{sec:tpcf}

The TPCF is a relatively straight forward way to study spatial structures. This method determines the distance be- tween all possible pairs of stars (d$_{s}$), shown as a histogram, which is then normalised to that of a reference distribution, i.e., N$_{\rm links}(d_{s}) / {\rm N}_{\rm reference}(d_{s})$. Gomez et al. (1993) used the TPCF to study the distribution of young stars in a star-forming region and used a random distribution for the reference distribution. This is not applicable in the current case as the galaxies have underlying structure, being centrally concentrated.  For the reference distribution we use an equal number of stars drawn from the red-giant branch, where we can be confident that each star has an age greater than $\sim1$~Gyr.

For the TPCF, an equal number of stars per age bin (and also for the reference distribution) is required. For the galaxies presented here, we use between 500 and 1250 stars per age bin, depending on the total number of BHeBs stars present. The exact number was chosen for each galaxy as a compromise in age resolution and having enough stars for the statistical analysis.  For all galaxies, multiple numbers of stars per age bin were used in order to check the robustness of the results. For each age bin, we generated five reference distributions, randomly sampling the same number of stars from the background population. The TPCF was then derived with each of the five reference distributions and the slope and y-offset of the resultant histogram was measured. The adopted values were the mean of the five distributions with the standard deviation being used to estimate the uncertainties.

Ideally, the slope and y-offset of the TPCF should converge to a value of zero if the population under study is evolving into the reference distribution. Deviations from zero imply that the reference population is not an ideal fit, where the offset is due to contamination or due to a real offset.  Differing distributions of young and old stars have been reported in the literature before (e.g., UGC~4438 - Dolphin et al.~2001).  However, since we will be studying {\it relative} changes, the adopted reference population is not critical.  

Since young stars are known to be more highly clustered relative to older stars, we expect the TPCF to have the largest offset from the reference population at young ages and evolve towards the reference distribution as older populations are considered. Ideally, we should see both the slope and the y-offset of the TPCF stop evolving at some characteristic time, t$_{\rm evo}$, where all of the original structure has been erased on galactic scales.  The measured TPCF for each galaxy in the sample is shown in the middle and bottom panels of the evolutionary figures (Figs.~\ref{fig:3panel_hoii}, \ref{fig:3panel_ddo165}, \ref{fig:3panel_ic2574}, \ref{fig:3panel_n2366}, \ref{fig:3panel_n784}, \& \ref{fig:3panel_n4068}).  Finally, in order to test for binning artefacts, we have also adopted smaller numbers of stars included in each age bin.  These bins, which have increased noise but higher time resolution, are shown as light filled squares in the evolutionary panels.

\subsection{$Q$-parameter}
\label{sec:q}

Cartwright \& Whitworth (2004; hereafter CW04) have developed a technique to measure the degree of substructure/degree of central concentration of a discrete two-dimensional dataset. We refer to the reader to Cartwright \& Whitworth (2004, 2009), Schmeja \& Klessen (2006), and B09 for a full description and tests of the technique.  The CW04 method, known as the $Q$ parameter, is based on Minimum Spanning Trees (MST), which for each dataset is formed by connecting all points (spatial positions in this case) in order to form a unified network, such that the total length (i.e., sum) of all of the connections, known as 'edges' or 'branches', is minimised, and no closed loops are formed. The mean branch length of the MST is calculated and then normalised by the factor $(N_{\rm total}A)^{0.5} / (N_{\rm total} - 1)$ to give $\bar{m}$, where $A$ is the area of the dataset and $N_{\rm total}$ is the total number of stars in the dataset.   The mean distance between all points (derived from the two-point correlation function) is then computed, which is then normalised by the radius of the dataset, giving $\bar{s}$.  In our case, the area is derived from the convex hull of the dataset, and the radius is assumed to be the radius corresponding to a circle with that area (see Schmeja \& Klessen 2006).  $Q$ is then simply $\bar{m}/\bar{s}$.

CW04 have carried out a series of numerical tests and have shown that $Q$ is highly sensitive to the amount of structure present (i.e., the degree of substructure) in a dataset.  Additionally, $Q$ can differentiate between a highly structured and a centrally concentrated distribution.  Their tests used fractal distributions to simulate clustering, where low fractal dimensions have large amounts of substructure, and high fractal dimensions are more uniformly distributed.  CW04 found that a 3D spherical distribution (projected onto a 2D surface) that is uniformly populated will have $Q=0.79$, while highly structured distributions will have $Q<0.79$ and centrally concentrated distributions will have $Q>0.79$.

In general, the lower the $Q$-value, the more substructure is present. As the dataset becomes more uniform or centrally concentrated, $Q$ increases. One complication is the effect of elongation on the value of $Q$.  B09 and Cartwright \& Whitworth~(2009) have shown that if a dataset is elongated, the $Q$ value will be lower than for the same spherical distribution.  This is important for the current work, given that many of the dIrrs in the current study are highly elongated and are seen from non-face-on orientations.  However, in the current work we shall not use the absolute value of the measured $Q$-value, but instead study the evolution of $Q$ as a function of age, which should be largely independent of elongation and other geometrical effects.

Additional tests on the effects of differential extinction and contamination on the measured $Q$-value have been discussed in detail in Appendix B and C of B09, respectively.   These are not expected to significantly affect the results presented in the current study.  We will focus our attention on relative changes in $Q$, hence (except for the youngest stellar populations) all populations should be similarly affected by extinction throughout the galaxy.

As with the TPCF, due to the highly structured distribution of young stars in galaxies, we expect Q to start with a low value and evolve towards the reference distribution and stop evolving once t$_{\rm evo}$ is reached. $t_{\rm evo}$ was estimated for each galaxy using the evolutionary figures (Figs.~\ref{fig:3panel_hoii}, \ref{fig:3panel_ddo165}, \ref{fig:3panel_ic2574}, \ref{fig:3panel_n2366}, \ref{fig:3panel_n784}, \& \ref{fig:3panel_n4068}) where the distributions stop changing.  This was done by eye, weighting the three measurements.

Tests of the stability of Q using the observed distribution as well as models showed that fewer sources are needed than for the TPCF method. An additional benefit of the Q-parameter method is that it is independent of the number of sources used. We take advantage of this feature by using smaller bins at young ages (where we expect the fastest and most pronounced structural evolution) and larger bins at older ages. Typically, between 500 and 750 stars were used in the first five age bins, and 1500 stars in the oldest bins. These points are shown as filled circles in the top panels of the evolutionary figures (Figs.~\ref{fig:3panel_hoii}, \ref{fig:3panel_ddo165}, \ref{fig:3panel_ic2574}, \ref{fig:3panel_n2366}, \ref{fig:3panel_n784}, \& \ref{fig:3panel_n4068}).  Since a constant number of sources is not necessary, we have also measured the Q-value for the OB stars in our sample. This is shown as a filled triangle in the evolutionary figures.

In order to test the reliability of the results, and any dependence on binning, we have also carried out the same analysis using smaller age bins.  While this increases the noise in the results, it can be used to confirm general trends.  These points are shown as light-shaded squares in the evolutionary figures.  Finally, we have estimated the uncertainty associated with the determination of the $Q$-value by picking N stars at random from the reference population of \hoii.  We find that the uncertainty in $Q$ is 0.027, 0.014, 0.011, 0.010 for $N=200, 500, 1000, \&~1500$, respectively (based on 100 random realisations of the background), which are shown in the figures as error bars.

\subsection{Caveats}
\label{sec:caveats}

While the two methods employed in the current work are powerful tracers of structural evolution, some caveats are necessary to discuss.  We have already discussed the need for using a representative reference distribution for the TPCF method (and to a smaller degree, the $Q$-method).  Below are a few additional caveats.

\begin{enumerate}

\item {\bf The existence of long lived structures (i.e., stellar clusters):}  Not all structures are expected to be destroyed within the first few hundred Myr of their lives.  Stellar clusters are gravitationally bound collections of stars that may live to very old ages.  However, only a very small fraction of stars in a galaxy are in clusters (after a few Myr -  e.g., Miller \& Scalo~1978; Adams \& Myers~2000; Goddard, Bastian, \& Kennicutt~2010).  Hence, stellar clusters are unlikely to significantly affect the results presented here.

\item {\bf Inclination effects:}  The galaxies in our sample are observed at various inclination angles.  While $Q$ has been developed in order to give a quantitative measure of the amount of structure present, the inclination angle and elongation of the system complicate the interpretation.  Hence, instead of absolute values, we will focus on relative changes within single galaxies.  In this way we can track the evolution of structures, but cannot quantify the fractal dimension of a population or the level of central concentration for the older populations.

\item {\bf Limited field of view:} In addition to inclination effects, some galaxies used in our survey have stellar populations that continue beyond the ACS coverage.  This also limits the use of the $Q$-value as an absolute measure.  However, the evolving spatial structure should be evident within the observed field-of-view of each galaxy, again meaning that we can track the relative change of structure.

\end{enumerate}

\section{Results}
\label{sec:results}

\subsection{Holmberg~{\sc II}}

Holmberg~{\sc II} is an actively star-forming dwarf galaxy in the M81 group and has an inclination angle of $i\sim44$~degrees (Puche et al.~1992).  Its current SFR (SFR$\sim0.17$~\msunyr) is roughly a factor of two higher than the average over the previous 500~Myr (W08), suggesting a recent increase in its activity.  The H{\sc I} distribution within \hoii\ is made up of numerous holes, although the distribution of red (presumably older) stars within the galaxy is quite smooth (W08).  The ACS observations used here do not cover the full extent of the galaxy, as both red and blue stars extend off the northwest of the field-of-view.  Due to the recent active star-formation within this galaxy, numerous high-mass (OB) stars are seen on the main-sequence (see Fig.~\ref{fig:cmd_hoii}).

Figure~\ref{fig:3panel_hoii} shows the evolution of the Q-value, TPCF slope, and TPCF y-offset for the BHeB stars (filled circles) and OB stars (filled triangle - only in the Q-value plot). Note the strong evolution in all indicators until an age of $225$~Myr. The distributions then level off, and remain approximately constant until age of $\sim600$~Myr (our last age bin). 

One important point to note is that the $Q$-value does not merge into the reference population nor does the TPCF converge to zero.  The reason for this is that the reference population (RGB stars) has a significantly different distribution than the BHeB stars at all ages.  This holds for all galaxies in our sample, meaning that we will limit our analysis to the relative change in each of the structural analysis methods. In the case of \hoii\ the RGB stars appear to be more centrally concentrated than the BHeB stars, which is opposite to what we see in all of the other galaxies in our sample.

\begin{figure*}
\includegraphics[width=16cm]{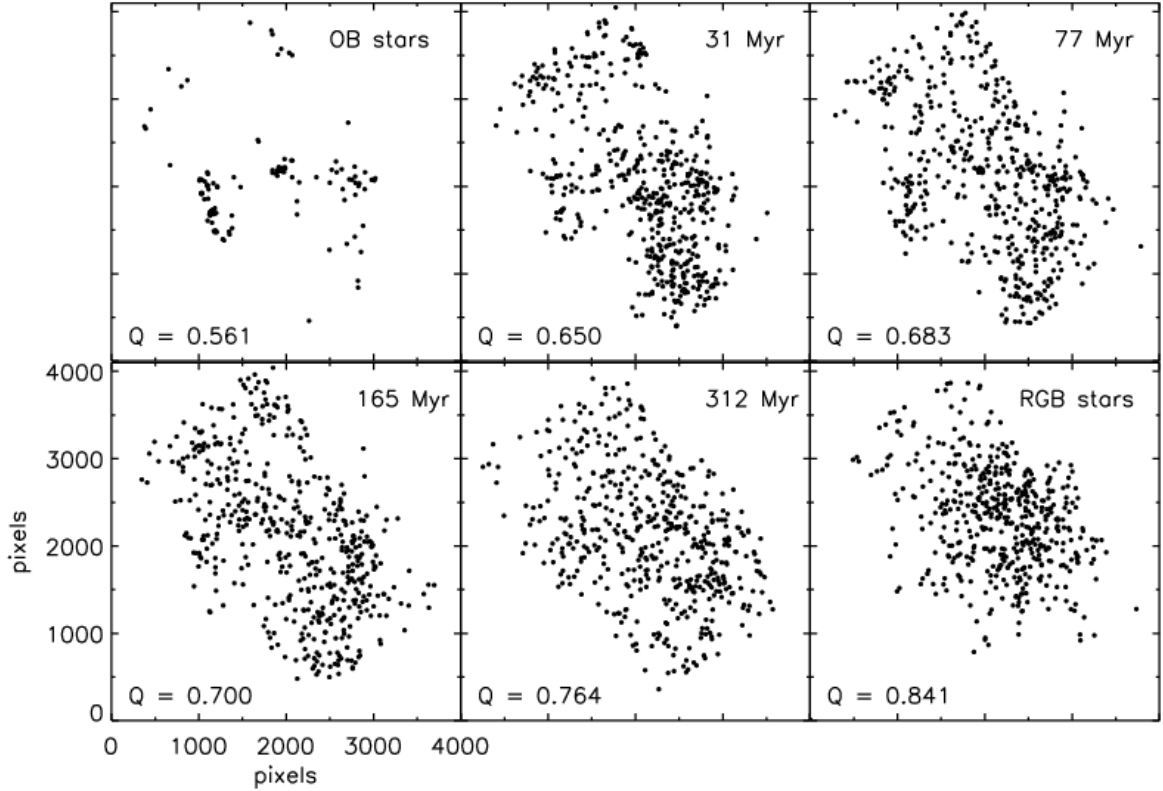}
\caption{The positions of the BHeB stars in \hoii\ for five age bins and a random selection from the reference RGB stars.  Each panel has 500 stars (except for the first, where we only have 126 OB stars) that were stochastically chosen from a given bin. The mean age of the bin and the measured $Q$-value is given in each panel.}
\label{fig:hoii_spatial}
\end{figure*}

\begin{figure}
\includegraphics[width=8.5cm]{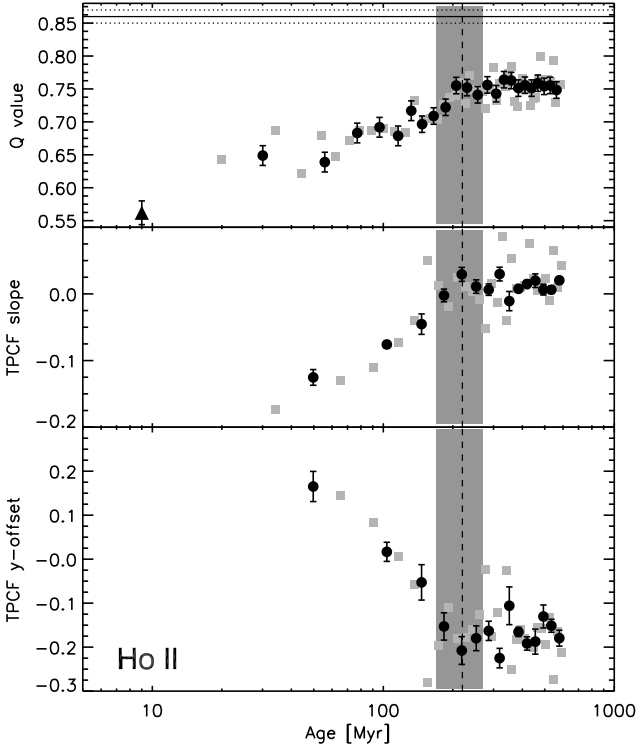}
\caption{The evolution of spatial structures in \hoii. {\bf Top panel:} The measured Q-value as a function of age. Filled circles represent the mean age and Q-value of BHeB stars while the filled triangle is that for OB stars selected from the CMD.  Lighter square symbols denote the results using more refined age bins, i.e. including fewer stars per bin.  The vertical dashed line is the adopted age (with uncertainty indicated as the filled area) where structural evolution stops.  The solid and dashed horizontal lines represent the value and error of the RGB background, respectively.  {\bf Middle panel:} The same as the top panel but now we show that evolution of the slope of the two-point correlation function (TPCF). {\bf Bottom panel:} The same as the middle panel, but now the y-offset of the TPCF. Note the strong evolution in Q and the TPCF until $\sim225$~Myr, followed by a flat distribution until an age of $\sim600$~Myr.}
\label{fig:3panel_hoii}
\end{figure}

In Fig.~\ref{fig:hoii_spatial} we show the spatial distribution of BHeB stars in \hoii\ for six age bins. The strong evolution in the clustering properties of the stars can be discerned by eye, which is quantified in Fig.~\ref{fig:3panel_hoii}.

\subsection{DDO~165}

\begin{figure}
\includegraphics[width=8.5cm]{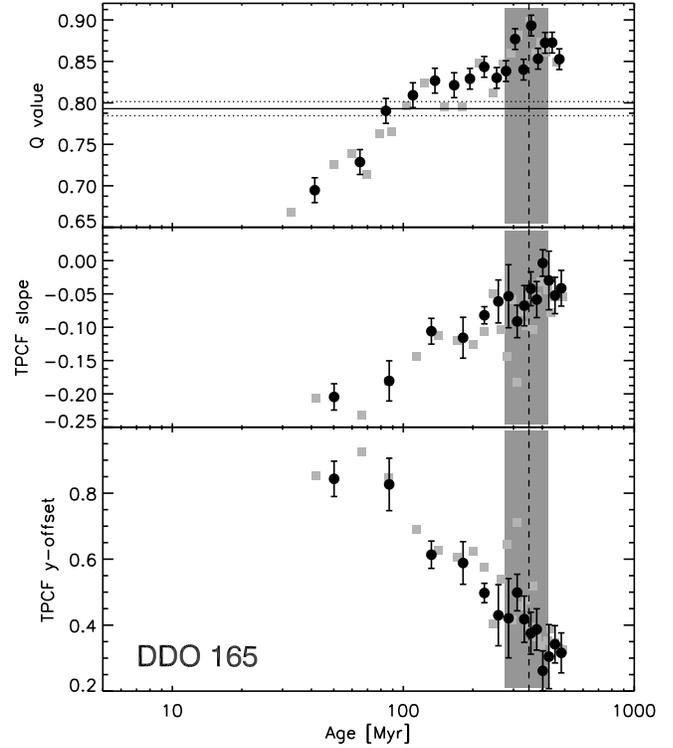}
\caption{The same as Fig.~\ref{fig:3panel_hoii} but now for DDO~165.}
\label{fig:3panel_ddo165}
\end{figure}

DDO~165, a dwarf galaxy in the M81 group, was covered by a single ACS pointing.  The CMD of DDO~165 shown in Fig.~\ref{fig:cmds} does not show a clear separation between BHeB stars and the main-sequence, showing just a single blue plume.  However, given the post-burst nature of this galaxy (Lee et al. 2007, W08), the MS is poorly populated at the brightest magnitudes compared to the BHeBs.  Brighter blue stars are more likely BHeBs, making identification trivial.  However, at fainter magnitudes, separation of the MS and BHeBs becomes more difficult.
The evolution of structure as traced by the BHeB stars can be seen in Fig.~\ref{fig:3panel_ddo165}. As with \hoii, we see a clear evolution in structure at young ages with all indicators. We adopt a value of $350\pm75$~Myr for \tevo\ for DDO~165, although we note that the observations are also consistent with a continuously evolving population until the oldest age bin ($\sim500$~Myr).

\subsection{IC~2574}

IC~2574 is a large, highly elongated, dwarf Irregular galaxy in the M81 group ($i\sim53$~degrees - Oh et al.~2008).  The galaxy has a ring of bright H{\sc ii} regions around an H{\sc i} hole  in the northeast of the galaxy, the brightest of which contains $\sim1/3$ of the ionising luminosity as 30~Doradus (Walter et al.~1998).  Three ACS fields were used to map its structure.  As seen in W08, the stellar distribution within IC~2574 is highly non-symmetric within the field-of-view, especially at young ages (blue stars).  These are concentrated largely in the northern-most ACS field, while the red stars are more symmetrically distributed (being most highly concentrated in the centre field-of-view).  IC~2574 appears to have undergone a period of intense star-formation, beginning approximately 1~Gyr ago and lasting for $\sim500$~Myr (W08).

\begin{figure}
\includegraphics[width=8.5cm]{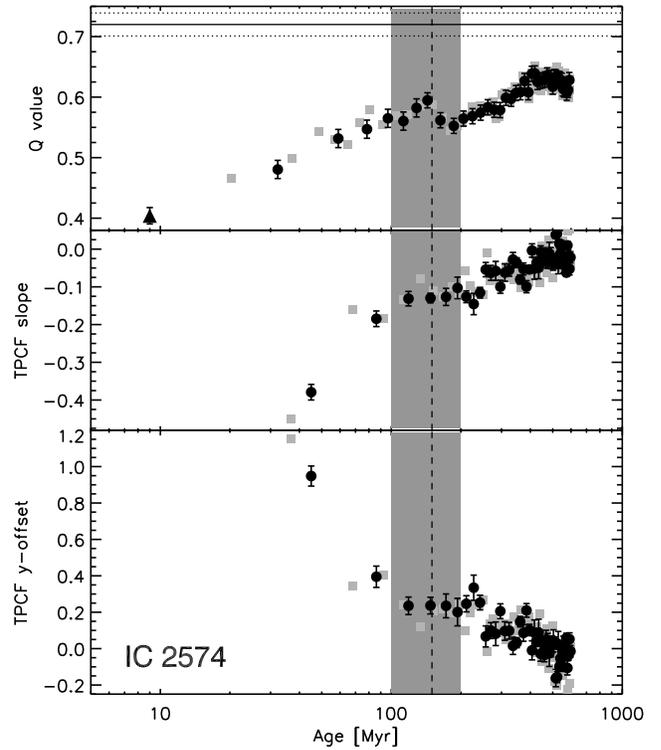}
\caption{The same as Fig.~\ref{fig:3panel_hoii} but now for IC~2574.}
\label{fig:3panel_ic2574}
\end{figure}

The evolution of the spatial structure in IC~2574 is shown in Fig.~\ref{fig:3panel_ic2574}.  In the upper panel, $Q$ evolves gradually from the youngest age bin (the OB stars) until $\sim150$~Myr, at which point it decreases until an age of $\sim200$~Myr at which point it begins to rise again until the oldest age bin. We see the same type of evolution in the TPCF method, although it is less apparent at young ages due to the higher number of sources needed in order to carry out the analysis (i.e., causing the age resolution to decrease).  We have used different age binning choices for the TPCF and find that the results corroborate the findings from the $Q$ evolution.

We conservatively adopt \tevo$=150$~Myr, although the structural properties of IC~2574 clearly continue to change until the oldest ages that we can study ($\sim500$~Myr).  Hence our value should be taken as a lower limit.

\subsection{NGC~2366}

NGC~2366 is the final dwarf galaxy in our sample from the M81 group.  Two ACS fields were required to observe the main optical extent of the galaxy.  The older stars show an elliptical, centrally concentrated structure that nearly fills the field-of-view, while the younger (blue) stars show a highly structured distribution (W08).  The galaxy contains a large H{\sc ii} region, NGC 2363 (Kennicutt et al.~1980) which contains a large population of young massive stars.  NGC~2366 appears to be undergoing a recent increase in its SFR, which began $\sim 50$~Myr ago (W08).  The inclination angle of the galaxy is $i\sim64$~degrees (Oh et al. 2008).

\begin{figure}
\includegraphics[width=8.5cm]{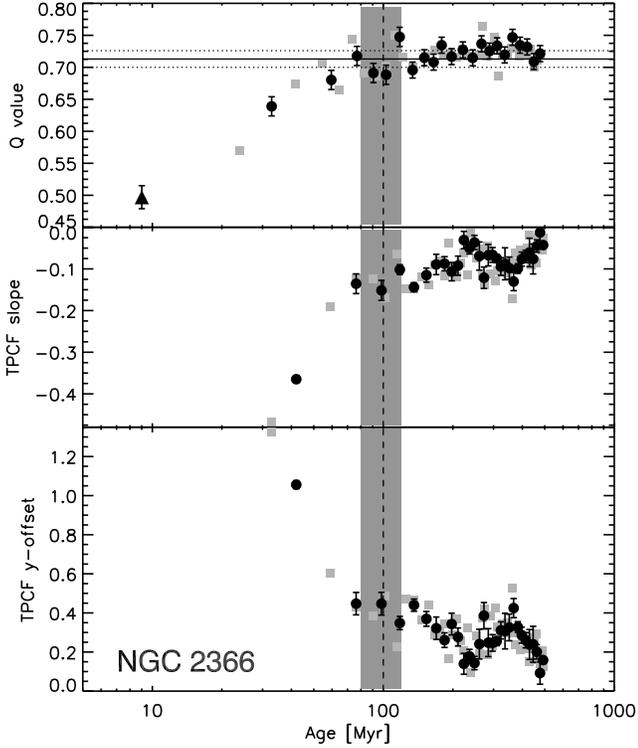}
\caption{The same as Fig.~\ref{fig:3panel_hoii} but now for NGC~2366.}
\label{fig:3panel_n2366}
\end{figure}

In Fig.~\ref{fig:3panel_n2366} we show the evolution of spatial structure within NGC~2366.  The large amounts of structure seen at young ages rapidly disappears.  From the evolution of $Q$, we see that the distribution becomes flat (i.e. lacks further change) at $\sim100$~Myr.  The TPCF slope and y-offset suggest a more complicated picture, changing until $\sim200$~Myr and then reversing the trend.  We do not have an explanation for this behaviour, and we adopt \tevo\ from the $Q$ evolution as a conservative lower limit.

\subsection{NGC~784}

NGC~784 is a barred dwarf Irregular galaxy seen nearly edge on.  The galaxy hosts a number of H{\sc ii} regions showing that star-formation is still actively ongoing (Kandalyan, Khassawneh, \& Kalloghlian~2003).  Due to its highly inclined position ($i\sim79$~degrees, Karachentsev et al.~2004), it is more difficult to see the evolution of structures by eye, although the TPCF and $Q$ evolution suggest that it is happening.

\begin{figure}
\includegraphics[width=8.5cm]{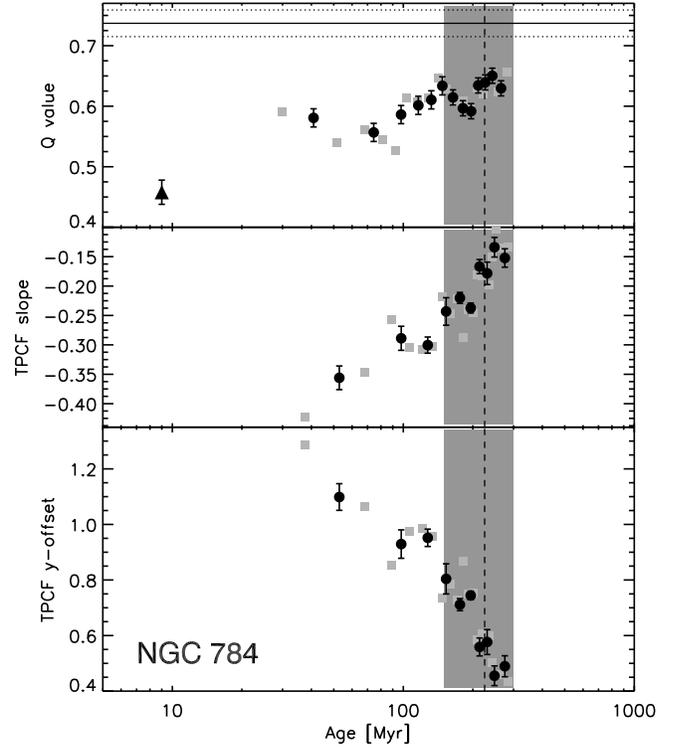}
\caption{The same as Fig.~\ref{fig:3panel_hoii} but now for NGC~784.}
\label{fig:3panel_n784}
\end{figure}

The TPCF and $Q$ evolution of BHeB stars, shown in Fig.~\ref{fig:3panel_n784} shows a smooth evolution from a highly structured distribution to a more uniform/centrally concentrated type grouping.  There is some evidence that the distribution stops evolving at an age of $225$~Myr, although it may continue past the last age bin available to us.  The young OB stars are consistent with the extrapolation of the $Q$-parameter from older ages.

\subsection{NGC~4068}

NGC~4068 is a blue compact dwarf (BCD) galaxy located at a distance of 4.2~Mpc, in the Canis Venatici I (CVn I) cloud (Karachentsev et al. 2006).  The galaxy is currently undergoing an increase in its star-formation activity, which began $\sim400$~Myr ago and is continuing until the present (McQuinn et al.~2009).  NGC~4068 has an inclination angle of $\sim60$ degrees (Karachentsev et al.~2004).

\begin{figure}
\includegraphics[width=8.5cm]{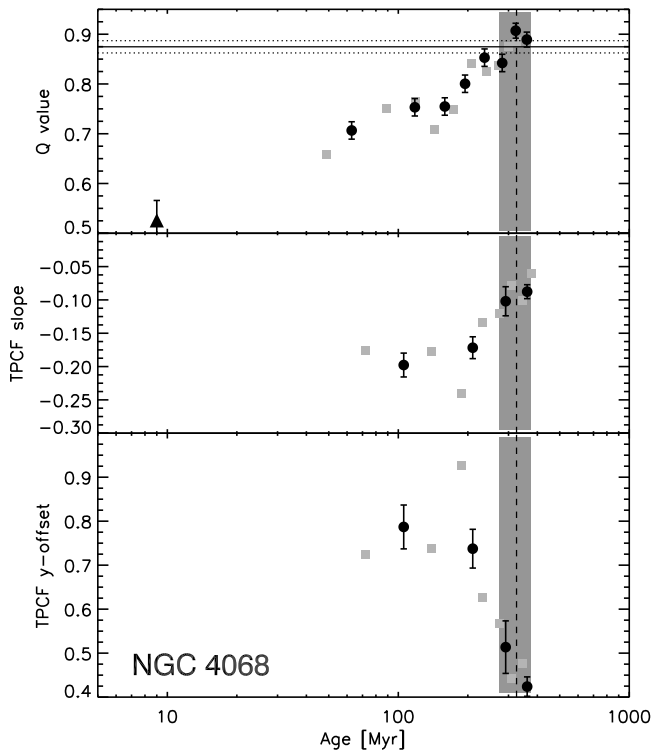}
\caption{The same as Fig.~\ref{fig:3panel_hoii} but now for NGC~4068.}
\label{fig:3panel_n4068}
\end{figure}

Figure~\ref{fig:3panel_n4068} shows the evolution of stellar structures in NGC~4068.  While the TPCF measurements are quite noisy, the $Q$ evolution appears to be quite smooth.  We assign a value of \tevo$\geq325\pm50$~Myr, although there is not strong evidence that the evolution of structure stops at this age.  While there are only a small amount of OB stars apparent in this galaxy (60), their position in the $Q$-diagram (top panel Fig.~\ref{fig:3panel_n4068}) is consistent with the extrapolation of the $Q$ evolution seen in the BHeBs.

\section{Discussion and Conclusions}
\label{sec:discussion}

Each galaxy in the present work shows clear evidence of evolving stellar structures.  Stars appear to be born in a highly structured, possibly hierarchical or fractal distribution, which evolves smoothly into a more uniform (centrally concentrated but lacking significant substructure) distribution.  Using two statistical methods, the $Q$ parameter and the Two Point Correlation Function (TPCF), we have been able to place lower limits on the timescale over which this transition occurs.  We find that this transition takes between $90$~Myr (NGC~2366) to $>300$~Myr (DDO~165).  These are similar timescales to those that we found in previous studies on the LMC ($175$~Myr - B09) and SMC ($75$~Myr - G08).    Hence, stellar structures remain coherent within these galaxies, on average, for over $200$~Myr.

The estimated evolutionary timescales, \tevo, are lower limits as we see the TPCF and $Q$-parameter continue to vary to some degree in all galaxies (with the exception of \hoii) until the oldest age bin.  Some galaxies, such as DDO~165, NGC~784, and NGC~4068 appear to have continuously evolving structures until the oldest ages under study.  Others, such as NGC~2366 and IC~2574 appear to show more of a composite evolution, with a rapid initial evolution ($100-150$~Myr), followed by either a period of no or opposite evolution, and finally a further decrease in substructure at large ages.  \hoii\ shows the cleanest evolution with a rapid decrease in structure lasting $\sim225$~Myr with no further evolution found until the oldest age bin in the sample ($\sim600$~Myr).

There does not appear to be a strong dependence on galaxy luminosity or galaxy size.  Unfortunately, stellar velocity dispersion measurements are not available for these galaxies.  For the LMC and SMC, where such measurements are available, B09 and G08 found that \tevo\ was comparable to the crossing time of the galaxy.  Such a prediction can be verified with the current study once the velocity dispersion of the young stellar components are measured.  Using the measured physical size of the galaxies and adopting the derived evolutionary timescales, we can estimate upper limits to the stellar velocity dispersions of these galaxies, assuming that structures erase on a crossing time.  As an example, our field of view of \hoii\ is approximately 360" by 200".  Approximating this area as a circle leads to a radius of $\sim150$" (or 2.46~kpc assuming a distance of 3.39~Mpc - Karachentsev et al.~2002).  Substituting \tevo\ for the crossing time ($t_{\rm cross} = R/\sigma$) leads to the upper limit of the velocity dispersion of $\sim11$~km\,s$^{-1}$.  As a substitute for the stellar velocity dispersion, we note that the           
mass-weighted median of the HI velocity dispersion in \hoii\ is  9.4 km s$^{-1}$ (Tamburro et al.~2009).  This may be taken as an upper bound for the velocity dispersion of the          
newly formed stars, as the measured velocity dispersions show both  cold (narrow) and broad (warm) components (see Young and Lo 1997 and discussion in Tamburro et al.~2009), and the star formation is  expected to be associated with the narrow component.  Thus, for \hoii\ there appears to be no discrepancy with the velocity dispersion  measurements and the existence of substructures corresponding to the crossing time.  For smaller galaxies, the crossing time argument becomes more restrictive, but, with these observations,  we do not see a trend in structure lifetime with galaxy size.                   

In nearly all cases, the RGB distribution (stars with ages $\gtrsim$1~Gyr) is significantly different than our oldest age bin accessible with the BHeB method, with the RGB stars being more extended.  This is consistent with previous findings, such as UGC~4438 (Dolphin et al.~2001) and NGC~6822 (de Blok \& Walter~2003) and similar to what we found for the LMC (B09) and SMC (G08).  Population studies and numerical modelling of galaxy formation and evolution (see Stinson et al.~2009) have shown that the outer parts of dIrrs are older on average than the inner parts.  At older ages, part of the evolution that we have detected in our sample of dIrrs may be due to this effect, and may not be directly attributable to dynamical structural evolution. The one counter-example in our sample is Holmerg~II, where the RGB stars appear to be more centrally concentrated than the BHeB stars.  Without considerably increasing the field-of-view of our observations, it is difficult to speculate why the distribution of RGB stars (relative to the younger BHeB stars) is different than for the other galaxies in our sample.

The main limitation of the current analysis is the lack of information on how stellar structures are dissolving on different spatial scales, as we have limited our study to the global distribution in each galaxy.  Studies that are able to probe the evolution of structure in both dimensions (in time and as a function of spatial scale) will be able to definitively test what mechanism is driving the observed evolution.

\section*{Acknowledgments}

We thank the referee for useful comments and suggestions.  Support for this work was provided by NASA through grant GO-10605 from the Space Telescope Science Institute, which is operated by AURA, Inc., under NASA contract NAS5-26555.  NB and BE acknowledge support from an STFC advanced fellowship and MG is supported by a Royal Society fellowship.

\appendix
\section{Spatial evolution of stellar structures for each galaxy considered}
\label{sec:app1}

Here we present examples of the spatial evolution of stellar structures for each galaxy in our sample.  Each figure is similar to Fig.~\ref{fig:hoii_spatial}.  In all cases, younger populations are seen to have much more substructure which dissolves in tens or hundreds of Myr.

\begin{figure*}
\includegraphics[width=16cm]{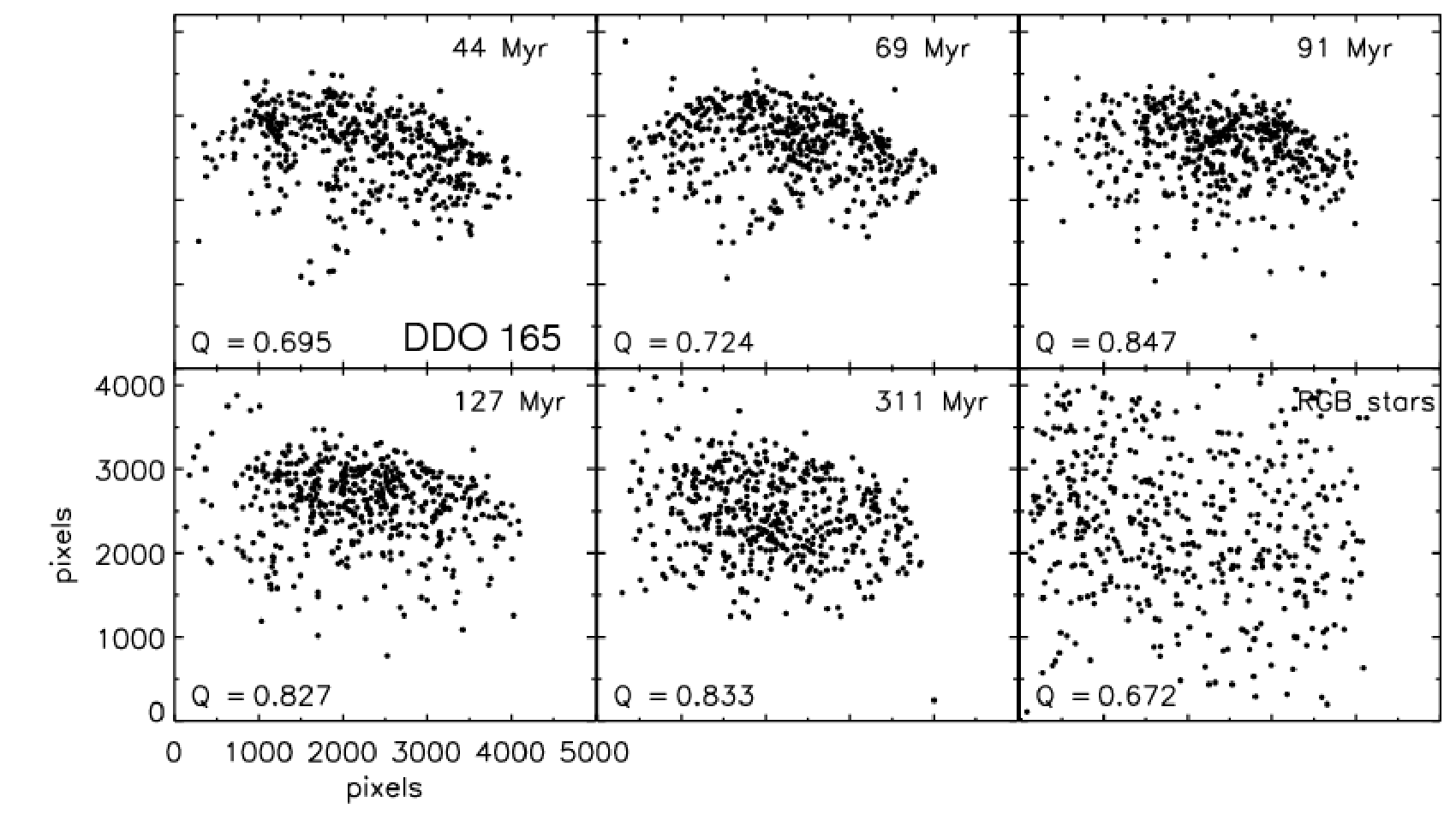}
\caption{The positions of the BHeB stars in DDO~165 (similar to Fig.~\ref{fig:hoii_spatial}) for five age bins and a random selection from the reference RGB stars.  Each panel has 500 stars  that were stochastically chosen from a given bin. The mean age of the bin and the measured $Q$-value is given in each panel.}
\label{fig:ddo165_spatial}
\end{figure*} 

\begin{figure*}
\includegraphics[width=16cm]{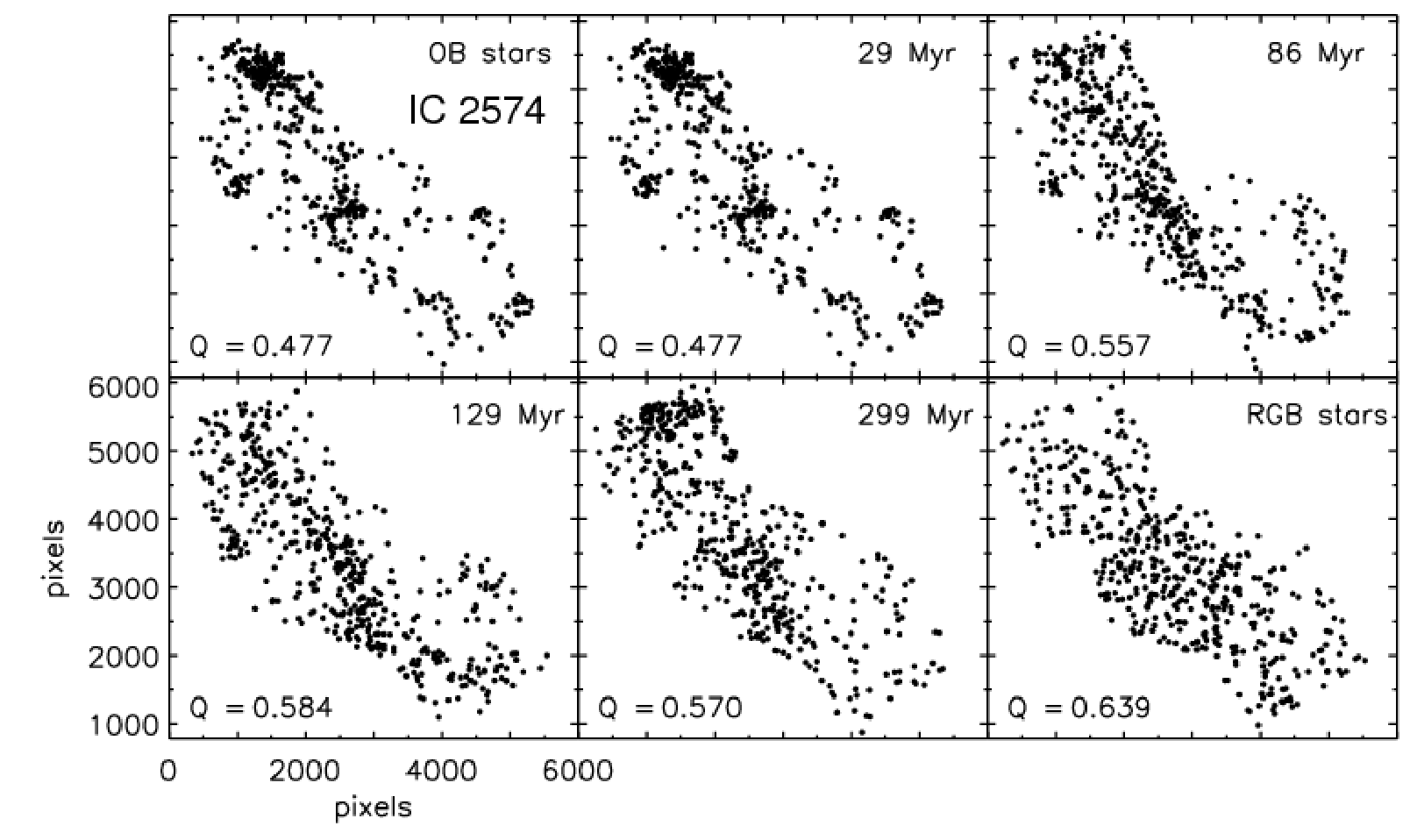}
\caption{Similar to Fig.~\ref{fig:hoii_spatial}, but now for IC~2547.}
\label{fig:ic2547_spatial}
\end{figure*} 

\begin{figure*}
\includegraphics[width=16cm]{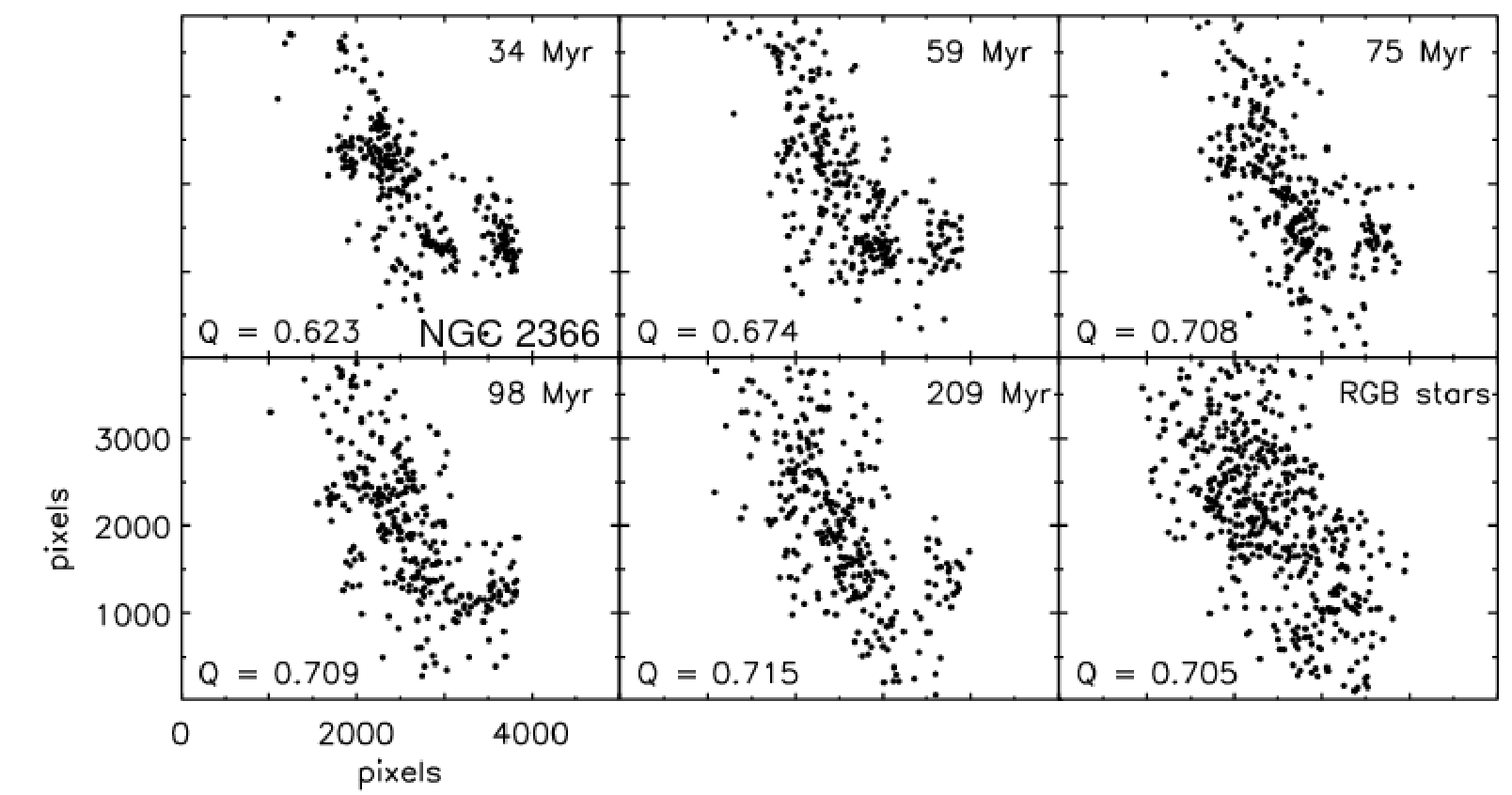}
\caption{Similar to Fig.~\ref{fig:hoii_spatial}, but now for NGC~2366.  Only 300 stars per age bin are shown in order to highlight the structural evolution at early times.}
\label{fig:n2366_spatial}
\end{figure*} 

\begin{figure*}
\includegraphics[width=16cm]{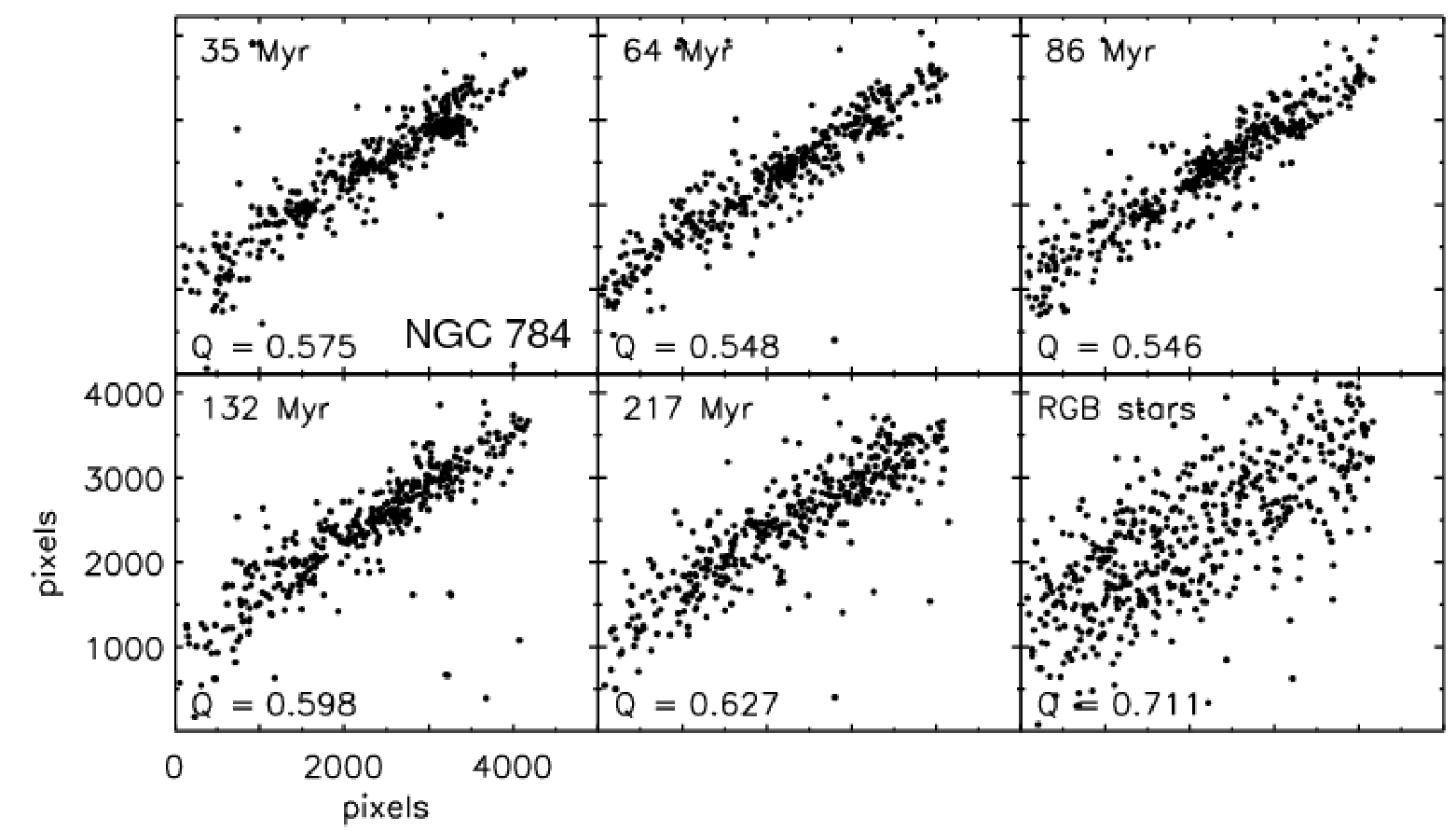}
\caption{Similar to Fig.~\ref{fig:hoii_spatial}, but now for NGC~784.  Only 400 stars per age bin are shown in order to highlight the structural evolution at early times.  Note that the edge on structure of the galaxy is likely to washout the the signal of structural evolution.}
\label{fig:n784_spatial}
\end{figure*} 

\begin{figure*}
\includegraphics[width=16cm]{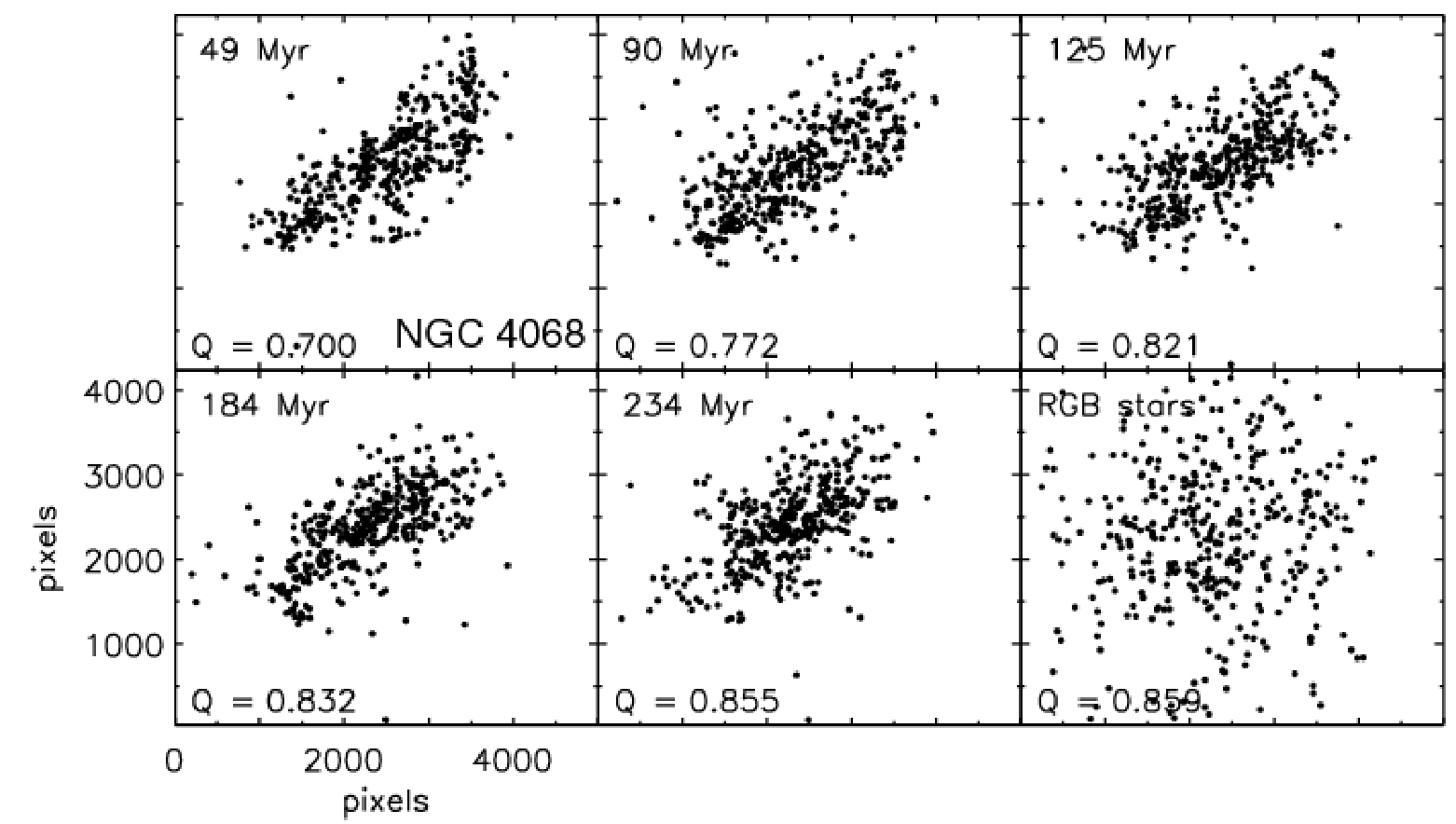}
\caption{Similar to Fig.~\ref{fig:hoii_spatial}, but now for NGC~4068.  Only 400 stars per age bin are shown in order to highlight the structural evolution at early times.}
\label{fig:n4068_spatial}
\end{figure*}

\bsp
\label{lastpage}

\begin{thebibliography}{99}
\bibitem[Adams 
\& Myers(2001)]{2001ApJ...553..744A} Adams, F.~C., \& Myers, P.~C.\ 2001, ApJ, 553, 744 
\bibitem[Bastian et 
al.(2005)]{2005A&A...443...79B} Bastian, N., Gieles, M., Efremov, Y.~N., \& Lamers, H.~J.~G.~L.~M.\ 2005, A\&A, 443, 79 
\bibitem[Bastian 
\& Goodwin(2006)]{2006MNRAS.369L...9B} Bastian, N., \& Goodwin, S.~P.\ 2006, MNRAS, 369, L9 
\bibitem[Bastian et al.(2007)]{2007MNRAS.379.1302B} Bastian, N., Ercolano, 
B., Gieles, M., Rosolowsky, E., Scheepmaker, R.~A., Gutermuth, R., 
\& Efremov, Y.\ 2007, MNRAS, 379, 1302
\bibitem[Bastian et al.(2009)]{2009MNRAS.392..868B} Bastian, N., Gieles, 
M., Ercolano, B., \& Gutermuth, R.\ 2009, MNRAS, 392, 868  
\bibitem[Bertelli et 
al.(1994)]{1994A&AS..106..275B} Bertelli, G., Bressan, A., Chiosi, C., Fagotto, F., \& Nasi, E.\ 1994, A\&AS, 106, 275 
\bibitem[Boily 
\& Kroupa(2003)]{2003MNRAS.338..665B} Boily, C.~M., \& Kroupa, P.\ 2003, MNRAS, 338, 665 
\bibitem[Cannon et al.(2003)]{can03} Cannon, J.~M., 
Dohm-Palmer, R.~C., Skillman, E.~D., Bomans, D.~J., C{\^o}t{\'e}, S., 
\& Miller, B.~W.\ 2003, AJ, 126, 2806
\bibitem[Cartwright  \& Whitworth(2004)]{2004MNRAS.348..589C} Cartwright, A., \& Whitworth, A.~P.\ 2004, MNRAS, 348, 589 
\bibitem[Cartwright 
\& Whitworth(2009)]{2009MNRAS.392..341C} Cartwright, A., \& Whitworth, A.~P.\ 2009, MNRAS, 392, 341 
\bibitem[Chiosi et al.(1992)]{chi92} Chiosi, C., Bertelli, G., \& Bressan, A.\ 1992, ARAA, 30, 235
\bibitem[Dalcanton et al.(2009)]{jdal09} Dalcanton, J.~J., et al.\ 2009, ApJS, 183, 67
\bibitem[de Blok 
\& Walter(2003)]{2003MNRAS.341L..39D} de Blok, W.~J.~G., \& Walter, F.\ 2003, MNRAS, 341, L39 
\bibitem[Dohm-Palmer et al.(1997)]{1997AJ....114.2527D} Dohm-Palmer, R.~C., et al.\ 1997, AJ, 114, 2527 
\bibitem[Dohm-Palmer et al.(1998)]{dom98} Dohm-Palmer, R.~C., et al.\ 1998, AJ, 116, 1227
\bibitem[Dohm-Palmer 
\& Skillman(2002)]{2002AJ....123.1433D} Dohm-Palmer, R.~C., \& Skillman, E.~D.\ 2002, AJ, 123, 1433 
\bibitem[Dohm-Palmer et al.(2002)]{2002AJ....123..813D} Dohm-Palmer, R.~C., Skillman, E.~D., Mateo, M., Saha, A., Dolphin, A., Tolstoy, E., Gallagher, 
J.~S., \& Cole, A.~A.\ 2002, AJ, 123, 813 
\bibitem[Dolphin(2000)]{2000PASP..112.1383D} Dolphin, A.~E.\ 2000, PASP, 112, 1383 
\bibitem[Dolphin(2002)]{2002MNRAS.332...91D} Dolphin, A.~E.\ 2002, MNRAS, 332, 91 
\bibitem[Dolphin et al.(2001)]{2001MNRAS.324..249D} Dolphin, A.~E., et al.~2001, MNRAS, 324, 249 
\bibitem[Goddard et al.(2010)]{2010MNRAS.tmp..491G} Goddard, Q.~E., 
Bastian, N., \& Kennicutt, R.~C.\ 2010, MNRAS, 491 
\bibitem[Goodwin \& Whitworth(2004)]{2004A&A...413..929G} Goodwin, S.~P., \& Whitworth, A.~P.\ 2004, A\&A, 413, 929 
\bibitem[Elmegreen 
\& Elmegreen(2001)]{2001AJ....121.1507E} Elmegreen, B.~G., \& Elmegreen, D.~M.\ 2001, AJ, 121, 1507 
\bibitem[Elmegreen et al.(2006)]{2006ApJ...644..879E} Elmegreen, B.~G., 
Elmegreen, D.~M., Chandar, R., Whitmore, B., 
\& Regan, M.\ 2006, ApJ, 644, 879 
\bibitem[Ford et al.(1998)]{for98} Ford, H.~C., et al.\ 1998, SPIE, 3356, 234
\bibitem[Gallart et al.(2005)]{gza05} Gallart, C., Zoccali, M., \& 
    Aparicio, A.\ 2005, ARAA, 43, 387 
\bibitem[Gieles et al.(2008)]{2008MNRAS.391L..93G} Gieles, M., Bastian, N., 
\& Ercolano, B.\ 2008, MNRAS, 391, L93 
\bibitem[Gutermuth et al.(2005)]{2005ApJ...632..397G} Gutermuth, R.~A., 
Megeath, S.~T., Pipher, J.~L., Williams, J.~P., Allen, L.~E., Myers, P.~C., 
\& Raines, S.~N.\ 2005, ApJ, 632, 397 
\bibitem[Gutermuth et al.(2008)]{2008ApJ...674..336G} Gutermuth, R.~A., et 
al.\ 2008, ApJ, 674, 336 
\bibitem[Kandalyan et al.(2003)]{2003Ap.....46...74K} Kandalyan, R.~A., 
Khassawneh, A.~M., \& Kalloghlian, A.~T.\ 2003, Astrophysics, 46, 74 
\bibitem[Karachentsev et al.(2004)]{2004AJ....127.2031K} Karachentsev,            
I.~D., Karachentseva, V.~E., Huchtmeier, W.~K., \& Makarov, D.~I.\ 2004, AJ, 127, 2031                                                                    
\bibitem[Karachentsev \& Kashibadze(2006)]{2006Ap.....49....3K} Karachentsev, I.~D., \& Kashibadze, O.~G.\ 2006, Astrophysics, 49, 3                                              
\bibitem[Karachentsev et al.(2006)]{2006AJ....131.1361K} Karachentsev, 
I.~D., et al.\ 2006, AJ, 131, 1361 
\bibitem[Koenig et al.(2008)]{2008ApJ...688.1142K} Koenig, X.~P., Allen, 
L.~E., Gutermuth, R.~A., Hora, J.~L., Brunt, C.~M., \& Muzerolle, J.\ 2008, ApJ, 688, 1142 
\bibitem[Kraus  \& Hillenbrand(2008)]{2008ApJ...686L.111K} Kraus, A.~L., \& Hillenbrand, L.~A.\ 2008, ApJL, 686, L111 
\bibitem[Lee et al.(2007)]{2007ApJ...671L.113L} Lee, J.~C., Kennicutt, R.~C., Funes, S.~J., Jos{\'e} G., Sakai, S., \& Akiyama, S.\ 2007, ApJL, 671, L113 
\bibitem[Marigo et al.(2008)]{mar08} Marigo, P., Girardi, L., Bressan, A., Groenewegen, M.~A.~T., Silva, L., \& Granato, G.~L.\ 2008, A\&A, 482, 883
\bibitem[McQuinn et al.(2010)]{2010ApJ...721..297M} McQuinn, K.~B.~W., et al.\ 2010, ApJ, 721, 297 
\bibitem[McQuinn et al.(2009)]{2009ApJ...695..561M} McQuinn, K.~B.~W., 
Skillman, E.~D., Cannon, J.~M., Dalcanton, J.~J., Dolphin, A., Stark, D., 
\& Weisz, D.\ 2009, ApJ, 695, 561
\bibitem[Miller \& Scalo(1978)]{1978PASP...90..506M} Miller, G.~E., \& Scalo, J.~M.\ 1978, PASP, 90, 506 
\bibitem[Odekon(2008)]{2008ApJ...681.1248O} Odekon, M.~C.\ 2008, ApJ, 681, 1248 
\bibitem[Oh et al.(2008)]{2008AJ....136.2761O} Oh, S.-H., de Blok, W.~J.~G., Walter, F., Brinks, E., \& Kennicutt, R.~C.\ 2008, AJ, 136, 2761 
\bibitem[Puche et al.(1992)]{1992AJ....103.1841P} Puche, D., Westpfahl, D., 
Brinks, E., \& Roy, J.-R.\ 1992, AJ, 103, 1841 
\bibitem[Sabbi et al.(2007)]{2007AJ....133...44S} Sabbi, E., et al.\ 2007, AJ, 133, 44 
\bibitem[Schmeja 
\& Klessen(2006)]{2006A&A...449..151S} Schmeja, S., \& Klessen, R.~S.\ 2006, A\&A, 449, 151  
\bibitem[Schmeja et al.(2008)]{2008MNRAS.389.1209S} Schmeja, S., Kumar, 
M.~S.~N., \& Ferreira, B.\ 2008, MNRAS, 389, 1209 
\bibitem[Schmeja et al.(2009)]{2009ApJ...694..367S} Schmeja, S., 
Gouliermis, D.~A., \& Klessen, R.~S.\ 2009, ApJ, 694, 367
\bibitem[Skillman et al.(1988)]{ski88} Skillman, E.~D., Terlevich, R., Teuben, P.~J., \& van Woerden, H.\ 1988, A\&A, 198, 33
\bibitem[Skillman(1996)]{ski96} Skillman, E.~D.\ 1996, The Minnesota Lectures on Extragalactic Neutral Hydrogen, 106, 208
\bibitem[Stinson et al.(2009)]{2009MNRAS.395.1455S} Stinson, G.~S.,  Dalcanton, J.~J., Quinn, T., Gogarten, S.~M., Kaufmann, T., 
\& Wadsley, J.\ 2009, MNRAS, 395, 1455 
\bibitem[Tamburro et al.(2009)]{2009AJ....137.4424T} Tamburro, D., Rix, H.-W., Leroy, A.~K., Mac Low, M.-M., Walter, F., Kennicutt, R.~C., Brinks,      
E., \& de Blok, W.~J.~G.\ 2009, AJ, 137, 4424                                  
\bibitem[Tully et al.(2006)]{tul6} Tully, R.~B., et al.\ 
2006, AJ, 132, 729
\bibitem[Walter et al.(1998)]{1998ApJ...502L.143W} Walter, F., Kerp, J., 
Duric, N., Brinks, E., \& Klein, U.\ 1998, ApJL, 502, L143 
\bibitem[Walter et al.(2007)]{2007ApJ...661..102W} Walter, F., et al.\ 
2007, ApJ, 661, 102 
\bibitem[Weisz et al.(2008)]{wei08} Weisz, D.~R., Skillman, E.~D., Cannon, J.~M., Dolphin, A.~E., Kennicutt, R.~C., Jr., Lee, J.C.,\& Walter, F.\ 2008, ApJ, 689,160
\bibitem[Young \& Lo(1997)]{1997ApJ...490..710Y} Young, L.~M., \& Lo, K.~Y.\ 1997, ApJ, 490, 710                                                                        


\end{thebibliography}
\end{document}